\documentclass[aps,prl,twocolumn,superscriptaddress,showpacs,showkeys]{revtex4}
\usepackage[centertags]{amsmath}
\usepackage{amsfonts}
\usepackage{amssymb}
\usepackage{amsthm}
\usepackage{newlfont}
\usepackage{graphicx}
\usepackage{hyperref}

\begin{document}


\title{A Case Study of Sedimentation of Charged Colloids: \\ The Primitive Model and the Effective One-Component Approach}



\author{Aldemar Torres}
\affiliation{Institute for Theoretical Physics, Utrecht University
\\ Leuvenlaan 4, 3584 CE Utrecht, The Netherlands.}

\author{Alejandro Cuetos}
\affiliation{Soft Condensed Matter Utrecht University.
\\ Princetonplein 5, 3584 CC Utrecht, The Netherlands.}

\author{Marjolein Dijkstra}
\affiliation{Soft Condensed Matter, Utrecht University.
\\ Princetonplein 5, 3584 CC Utrecht, The Netherlands.}

\author{Ren\'e van Roij}
\affiliation{Institute for Theoretical Physics, Utrecht University
\\ Leuvenlaan 4, 3584 CE Utrecht, The Netherlands.}


\date{\today}

\begin{abstract}

Sedimentation-diffusion equilibrium density profiles of
suspensions of charge-stabilized colloids are calculated
theoretically and by Monte Carlo simulation, both for a
one-component model of colloidal particles interacting through
pairwise screened-Coulomb repulsions and for a three-component
model of colloids, cations, and anions with unscreened-Coulomb
interactions. We focus on a state point for which experimental
measurements are available [C.P. Royall et al., J. Phys.: Cond.
Matt. {\bf 17}, 2315 (2005)]. Despite the apparently different
picture that emerges from the one- and three-component model
(repelling colloids pushing each other to high altitude in the
former, versus a self-generated electric field that pushes the
colloids up in the latter), we find similar colloidal density
profiles for both models from theory as well as simulation,
thereby suggesting that these pictures represent different view
points of the same phenomenon. The sedimentation profiles obtained
from an effective one-component model by MC simulations and
theory, together with MC simulations of the multi-component
primitive model are consistent among themselves, but differ
quantitatively from the results of a theoretical multi-component
description at the Poisson-Boltzmann level. We find that for small
and moderate colloid charge the Poisson-Boltzmann theory gives
profiles in excellent agreement with the effective one-component
theory if a smaller effective charge is used. We attribute this
discrepancy to the poor treatment of correlations in the
Poisson-Boltzmann theory.

\end{abstract}


\maketitle

\section{\label{sec:intro}Introduction}
Colloidal particles with a density different from that of the dispersive medium sediment because of the gravitational force. At fixed temperature
$T$, the resulting non-homogeneous equilibrium distribution is a consequence of the balance between energy and entropy of the different chemical
species involved. This equilibrium is characterized by measurable density profiles \cite{Piazza,Experiments}. In the case of sufficiently dilute
suspensions those profiles obey the barometric law $\rho(x)\propto{\exp(-x/L)}$, with $L=k_BT/mg$ the gravitational length, $m$ the buoyant mass,
$g$ the gravitational acceleration, $k_B$ the Boltzmann constant, $T$ the absolute temperature, and $\rho(x)$ the number density of colloids at
altitude $x$. In dense systems, with non-negligible colloidal interactions, strong deviations from the barometric law have been observed, e.g. for
colloidal hard-spheres at packing fractions up to and beyond the freezing point \cite{Piazza,Rutgers}. More surprisingly (at least initially) were
the strong deviations from the barometric law in rather dilute suspensions of highly charged colloids at low salinity \cite{Philipse exp}. The
measured density profiles suggested an extreme enhancement, by at least one order of magnitude, of the apparent mass of the colloids
\cite{Philipse exp,Experiments}. This system was theoretically analyzed in terms of a three-component model of colloids and monovalent cations and
anions, for which Poisson-Boltzmann (PB) theory in gravity revealed a self-consistent electric field that pushes up the colloids to high altitude
against gravity, thereby reducing the apparent mass as observed experimentally \cite{Rene's theory}. However, another explanation was given more
recently in Ref. \cite{belloni}, where hydrostatic equilibrium in a one-component system of colloids interacting through pairwise screened-Coulomb
repulsion was considered. In the present paper we investigate the relations between these two pictures in more detail by considering both models
(colloid-cation-anion mixture and colloids-only system) by theory as well as Monte Carlo simulation.  For the one-component approach we use a
model based on effective pairwise screened-Coulomb interactions. The profiles are obtained from the solution of the hydrostatic equilibrium
equation, that uses the isothermal compressibility obtained from the solution of the Ornstein-Zernike (OZ) equation with the rescaled mean
spherical approximation (RMSA) closure \cite{Rescaled MSA}. This approach is similar to that introduced in \cite{belloni}. In this case, entropic
and electrostatic effects are implicitly included in the structure of the suspension (see next section). Furthermore, we perform MC simulations
for this model. Within the multi-component picture, we approach the problem using the Poisson-Boltzmann theory introduced in Ref. \cite{Rene's
theory}, which explains the non-barometric profiles in terms of a macroscopic electric field that appears as a consequence of a charge
inhomogeneity. We also performed simulations of this system using the primitive model in gravity. These particular simulations require a
substantial amount of CPU time since a considerable number of micro-ions has to be taken into account to mimic the substantial salt concentration
in the suspension. We focus on a state point for which experimental information is available, and compare the profiles obtained from the mentioned
theoretical and numerical approaches with published measurements \cite{Experiments}.

The paper is organized as follows. In section 2 we introduce the one-component model, the structure factor of the suspension determined through
RMSA, and we discuss some details regarding the simulation technique of the one-component model. In section 3 we briefly revise the PB theory for
sedimentation, introduce the primitive model in gravity and some aspects regarding the simulations. Results and discussion are presented in
section 4 and 5 respectively, where we compare the different theoretical and numerical results with experimental data. A summary and conclusions
are gathered in section 6.

\section{Effective One-Component Model}

 Let us consider a system consisting of charged spheres of diameter $\sigma=2a$, mass $M$, and
electric charge $-Ze$, in osmotic contact with a reservoir of 1:1 electrolyte with salt concentration $2\rho_s$. The solvent has mass density
$\rho_l$ and is characterized by an electric permittivity $\epsilon$. Let us assume also that the dielectric constant of the spheres and the
electrolyte are identical to avoid electrostatic image effects and Van der Waals forces between the spheres. Assuming pairwise effective colloidal
interactions, the Hamiltonian of the effective one-component system of colloids in the presence of gravity is given by
\begin{equation}
\label{Hamiltonian One}
\textit{H}=\sum\limits_{i=1}^{N}mgx_i+\sum\limits_{i<j}^{N}v(r_{ij}),
\end{equation}
where the first term in the right-hand side is the potential
energy of colloid $i$ at height $x_i$ measured from the bottom of
the sample. Here $m=M-\rho_l\pi\sigma^3/6$ the buoyant mass of the
colloidal particles, and $v(r)$ is the familiar screened-Coulomb
potential
\begin{equation}
  \label{DLVO Potential}
\beta{v(r)} =
\begin{cases}
\displaystyle { \infty}\,,
&\text{if\ } {r}<\sigma,\\
\displaystyle{\frac{Z^2\exp({\kappa\sigma})}{(1+\kappa{a})^2}\frac{\lambda_B}{r}\exp{(-\kappa{r})}}
\,, & \text{if\ }{r}\geq\sigma,
\end{cases}
\end{equation}
with $\beta=(k_BT)^{-1}$ where $k_B$ is the Boltzmann constant and
where $\lambda_{B}=\frac{{\beta}e^{2}}{\epsilon}$ is the Bjerrum
length, $\kappa=\sqrt{8\pi\lambda_B\rho_s}$ is the inverse
screening length, and $r$ is the distance between centers of
colloidal particles. Under isothermal conditions and for small
density gradients, the osmotic pressure of the suspension depends
only on the local number density of colloids $\rho(x)$. The latter
is determined from the non-linear differential equation that
follows from inserting $\rho(x)$ into the hydrostatic equilibrium
equation $d\Pi(x)/dx=-mg\rho(x)$ with $\Pi$ the osmotic pressure
of the suspension with respect to the salt reservoir. This yields
\begin{equation}
\label{hydro}
\frac{d\rho(x)}{dx}+\frac{\chi_T(\rho(x))}{L}\rho(x)=0,
\end{equation}
where $\chi_T^{-1}=\left(
{\frac{\partial(\beta\Pi)}{\partial\rho}}\right)_{T}$ is the
isothermal compressibility of the bulk fluid and $L$ is the
gravitational length defined above.  The sedimentation profiles
can be obtained by solving (\ref{hydro}) if the function
$\chi_T(\rho)$ is known for the relevant density regime. In order
to determine $\chi_T(\rho)$ we use the well-known Kirkwood-Buff
relation $\chi_T=\lim\limits_{q\rightarrow 0}S(q)$ with $S(q)$ the
structure factor as calculated by Hansen, Hayter and Penfold
within the RMSA closure of the Ornstein-Zernike equation
\cite{Hayter Penford,Rescaled MSA}. By this procedure the
sedimentation profiles are determined solely from the structure of
the effective one-component bulk fluid. Notice that such a scheme
was applied successfully to explain the measured hard-sphere
density profiles in Ref. \cite{Piazza}. For later comparison we
also consider an alternative expression for $\chi_T(\rho)$, that
is based on the Donnan equation of state as e.g. given in Ref.
\cite{Rene's theory}. This yields
\begin{equation}
\label{Donnan}
\chi_T^{-1}=1+\frac{Z^2\rho/2\rho_s}{\sqrt{1+(Z\rho/2\rho_s)^2}},
\end{equation}
which features the high-density or low-salt limit
$\chi_T(\rho)=1+Z$ for $Z\rho\gg2\rho_s$, such that insertion into
(\ref{hydro}) yields $\rho(x)\propto\exp[-x/(Z+1)L]$, i.e. an
effective gravitational length that is a factor $Z+1$ larger than
that in the barometric law \cite{Rene's theory}. The remaining
task in order to find the sedimentation profiles is to insert
$\chi_T$ into (\ref{hydro}) and to solve the non-linear equation
numerically on an $x$-grid.

In addition we perform standard Monte Carlo simulations of a
system described by the interaction Hamiltonian (\ref{Hamiltonian
One}) for the parameters $Z=76$, colloid diameter
$\sigma=1.91\mu$m, Bjerrum length $\lambda_B=10.4$ nm, screening
parameter $\kappa\sigma=1.2$ and average colloidal packing
fraction $\overline{\eta}=H^{-1}\int_0^H\eta(x)dx=0.0053$ with the
height $H=50.92\sigma$. The experimental screening parameter
satisfies $\kappa\sigma=1.2$. These parameters are identical to
those of the experimental system studied in
Ref.\cite{Experiments}, where $Z=76$ stems from the best fit of
the experimental density profile with a theoretical prediction
based on the primitive model (see below). The dimensions of the
rectangular simulation box are $10\sigma\times 10\sigma\times{H}$.
We checked that the horizontal area was large enough to exclude
finite-size effects. We employed periodic boundary conditions in
the horizontal directions, in the vertical directions the system
is bounded by hard walls that exclude the centers of colloids at
$x<0$ and $x>H$.

\section{The Primitive Model in Gravity}

As mentioned in the introduction, a different approach to study
sedimentation profiles is to consider each chemical species
separately, namely colloids (c), coions (-) and counterions (+).
The Hamiltonian of the system can be written as
\begin{equation} \label{H primitive}
 \textit{H}=H_{cc}+H_{ii}+H_{ci}+\sum\limits_{i=1}^{N}mgx_i,
\end{equation}
where the first three terms in the right-hand side include
colloid-colloid, ion-ion and colloid-ion pairwise interactions,
respectively, and the last term is the gravitational energy of the
colloids introduced in equation (\ref{Hamiltonian One}); the ions
are assumed to be massless. In this model, the electrostatic pair
interactions are of the form
\begin{equation}
 \label{primitive}
\beta{v_{ij}(r)} =
\begin{cases}
\displaystyle { \infty}\,,
&\text{if\ } {r}<\sigma_{ij}=(\sigma_i+\sigma_j)/2,\\
Z_iZ_j\frac{\lambda_B}{r} \,, & \text{if\ } {r}\ge \sigma_{ij},
\end{cases}
\end{equation}
with $\sigma_k$ and $Z_k$ the diameter and the charge number of
species $k=\{c,+,-\}$, i.e. $Z_c=-Z,Z_+=1,Z_-=-1$,
$\sigma_c=\sigma$ and $\sigma_+=\sigma_-\ll\sigma$. The number of
particles are denoted by $N_k$, i.e $N_c=N$ and $N_+=N_-+ZN$ for
charge neutrality reasons. This three-component model can be
studied within Poisson-Boltzmann theory \cite{Rene's theory},
which relates the density profiles $\rho(x),\rho_+(x),\rho_-(x)$
of the colloids cations and anions respectively, to the local
electrostatic Donnan potential $\psi(x)$ through
\begin{eqnarray}
&&\rho _{\pm }(x)=\rho _{s}\exp [{\pm \phi (x)];}  \label{PB1} \\
&&\rho (x)=\rho _{0}\exp [{-x/L+Z\phi (x)];}  \label{PB2} \\
&&\phi ^{\prime \prime }(x)=\kappa ^{2}\sinh \phi (x)+4\pi \lambda
_{B}Z\rho (x),  \label{PB3}
\end{eqnarray}
with $\phi(x)$ defined by the dimensionless combination
$\phi(x)=e\psi(x)/k_BT$ and where $\rho_0$ is a normalization
constant. Here a prime denotes a derivative w.r.t. $x$. Under
appropriate conditions, typically $Z^2\rho(x)\gg2\rho_s$, it was
found that $\phi(x)$ is a linear function of $x$ in
macroscopically large parts of the system, i.e. there is a
constant electric field that lifts the colloids to higher
altitudes than expected on the basis of their mass \cite{Rene's
theory}. This result stems both from numerical solutions of
(\ref{PB1})-(\ref{PB3}) and from the "Laplace-Boltzmann" equation,
where (\ref{PB3}) is replaced by the local charge neutrality
condition $\kappa ^{2}\sinh \phi (x)+4\pi \lambda _{B}Z\rho
(x)=0$. Note that by combining the latter equation with
(\ref{PB1}) and (\ref{PB2}) one recovers the hydrostatic
equilibrium condition (\ref{hydro}) with $\chi_T$ given by
(\ref{Donnan}). On the other hand, the set of equations
(\ref{PB1})-(\ref{PB3}) can be solved numerically in order to
determine the local electrostatic potential $\phi(x)$ together
with the equilibrium profile $\rho(x)$ by an iterative procedure
as pointed out in \cite{Rene's theory}. Such a procedure requires
two boundary conditions, e.g. $\phi'(0)=\phi'(H)=0$, where $H$ is
the height of the solvent meniscus.

A system described by the Hamiltonian (\ref{H primitive}) was
simulated in a rectangular box of horizontal area $9\sigma\times
9\sigma$ and height $H=50\sigma$. The vertical coordinate $x$ is
restricted to $x\in[0,H]$, and periodic boundary conditions are
only applied in the horizontal plane and not in the vertical
direction. In order to be as closely as possible to the
experiments of Ref.\cite{Experiments}, we considered colloids with
charge $Z=76$, diameter $\sigma=1.910\mu$m, gravitational length
$L=2.41\sigma$, Bjerrum length $\lambda_B=10.4$nm, and average
colloidal packing fraction
$\overline{\eta}=H^{-1}\int_0^H\eta(x)dx=0.0053$ (with
$H=50\sigma$). The experimental screening parameter satisfies
$\kappa\sigma=1.2$. This state point is realized, with the present
box size and shape, by the number of colloids $N_c$=12 and the
number of positive and negative ions $N_+=$13516 and $N_-=$12604,
respectively. In order to take  the long-range electrostatic
interactions into account we have employed a combination of Ewald
summation in a slab geometry with the lattice method proposed by
Panagiotopoulos and Kumar \cite{PAN99}. The parameters of the
Ewald summation and lattice method are the same as those in
reference \cite{CUE06}. Note that the large number of ions, which
have to be included to represent the low but yet substantial
screening parameter, makes the simulations extremely time
consuming. A typical simulation consists of $10^5$ MC cycles. A
cycle consists of $0.9\mathcal{N}$ trials to move a randomly
chosen colloid and $0.1\mathcal{N}$ trials to move a randomly
chosen colloid or microion, with $\mathcal{N}=N+N_++N_-$ the total
number of microions and colloids in the system. In a dense system
of microions, a simple Monte Carlo move of a colloid would almost
certainly result in an hard-core overlap with one of the
microions. In order to avoid such overlaps we use a cluster move
technique, where ions that overlap with the new colloid position
are moved into the space left empty by the displaced colloid, more
details on this technique can be found in Ref. \cite{LOB03,HYN05}.
The percentage of accepted moves of each component (colloids and
microions) was maintained at about $40\%$. To check if the system
was equilibrated, the average altitude of the centers of mass of
the colloids was monitored in the simulation; when the center of
mass was not stable, further equilibration was performed before
taking measurements. A final simulation with $2\cdot10^5$ MC
cycles was performed to obtain averages.

\section{Results}

\begin{figure}
\label{compress}
\includegraphics[width=60mm,angle=-90]{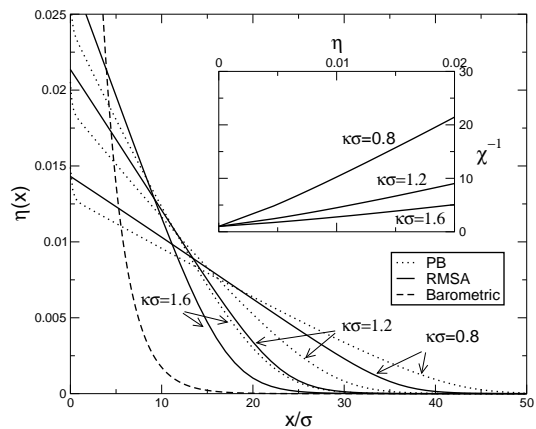}
\caption{Colloidal sedimentation profiles based on hydrostatic equilibrium (\ref{hydro}) calculated using the RMSA-based compressibility of the
one-component Yukawa model, compared to those based on the multi-component PB theory (\ref{PB1})-(\ref{PB3}), for the screening parameters
$\kappa\sigma=0.8, 1.2$ and 1.6. The colloidal charge is $Z=76$, the colloid diameter $\sigma=1.91\mu$m, the Bjerrum length is $\lambda_B=10.4$
nm, the gravitational length $L=2.41\sigma$, the average colloid packing fraction is $\overline{\eta}=0.0053$, and the sample height is
$H=50\sigma$, as reported in Ref. \cite{Experiments}. The dashed curve is the barometric distribution with the same normalization. The inset shows
the RMSA-based compressibility.}
\end{figure}

Figure 1 shows a first comparison of theoretical predictions based
on the one-component and three-component model. We see several
theoretical sedimentation profiles as a function of the altitude
$x$, corresponding to colloids of diameter $\sigma=1.91\mu$m, and
three different salt concentrations characterized by screening
parameters $\kappa\sigma=0.8$, 1.2, and 1.6. All profiles shown in
Fig.1 are for the same gravitational length $L=2.41\sigma$,
average packing fraction $\overline{\eta}=0.0053$, sample height
$H=50\sigma$, and Bjerrum length $\lambda_B=10.4$nm. For each
$\kappa\sigma$, the colloid density profile is calculated for both
the effective one-component model based on the solution of the
hydrostatic equilibrium equation (\ref{hydro}) using the
isothermal compressibility obtained from the RMSA closure, as well
as from the multi-component PB theory described by equations
(\ref{PB1})-(\ref{PB3}). We also show, for the sake of comparison,
the corresponding barometric profile obtained from (\ref{PB2}) and
(\ref{PB3}) in the case of uncharged colloids ($Z=0$), for the
same normalization. The inset shows the compressibilities as a
function of the colloid density, as obtained from the solution of
the OZ equation within the RMSA closure. At zero density all the
compressibility curves reduce to the ideal-gas compressibility,
and with increasing colloid density the electrostatic repulsions
manifest themselves as a reduction of $\chi_T$: the weaker the
screening, the stronger the effective colloidal interactions.  We
note that each one-component Yukawa system yields steeper density
profiles than those of the corresponding three-component model,
for all $\kappa\sigma$ considered here, i.e. the one-component
systems have a relatively small average altitude and a relatively
low density at higher altitudes. We will argue in more detail
below that the source of the difference between the one- and
three-component predictions is mainly due to the poor
representation of the colloid-colloid correlations in the
three-component PB theory.

\begin{figure}
\label{sim}
\includegraphics[width=70mm,angle=-90]{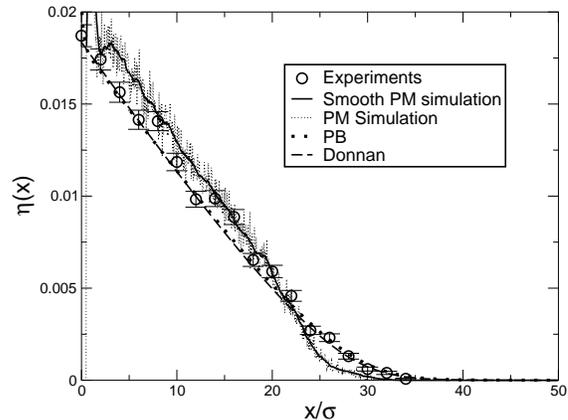}
\caption{Colloidal sedimentation density profiles for colloidal charge $Z=76$ stemming from multi-component PB theory (\ref{PB1})-(\ref{PB3}),
from the Donnan compressibility (\ref{Donnan}) combined with hydrostatic equilibrium (\ref{hydro}), and simulations of the Primitive Model in
gravity, for the parameters of Fig.1 and $\kappa\sigma=1.2$. The symbols denote the experimental measurements from Ref.\cite{Experiments}.}
\end{figure}

Figure 2 shows experimentally measured density profile of Ref.
\cite{Experiments} compared to density profiles as obtained from
the multi-component models: the three-component PB theory of
equations (\ref{PB1})-(\ref{PB3}) and simulation of the primitive
model in gravity as introduced in section 3. In spite of the fact
that the primitive model simulation was equilibrated during about
one year CPU time and the measurements were performed over four
months CPU time, the level of noise in the raw data is still quite
high. Therefore also a smoothed curve of the simulated profile is
shown to facilitate comparisons. The difficulty in obtaining good
statistics in this particular simulation is due to the fact that
the total number of colloids in the system is exceedingly small
($N_c=12$), whereas the total number of particles in the system is
rather large (26132), most of them salt ions needed to achieve the
required screening parameter condition. That the Donnan-based
density profile is accurate when compared to the experiments is
\emph{only} due to the fact that the experimental value $Z=76$
\footnote{The measured density profile was fitted to the
predictions of Poisson-Boltzmann theory, and it was concluded in
Ref. \cite{Experiments} that the colloidal charge equals $-78e$.
Here we concluded that $-76e$ gives the best fit within
Poisson-Boltzmann theory, this difference is due to details of the
fitting procedure and does not interfere with our arguments.}
stems from a fit to PB theory \cite{Experiments}, which is
equivalent to the Donnan equation of state in the local neutrality
"Laplace-Boltzmann" limit as explained below
Eqs.(\ref{PB1})-(\ref{PB3}). In other words, $Z=76$ is close to a
best fit to the Donnan equation of state.

In figure 3 we see a first comparison of the sedimentation profile
obtained experimentally with sedimentation profiles calculated
using the effective one-component models: simulation of the Yukawa
system and the RMSA approach of section 2. We also include the
profiles obtained from the multi-component PB theory for the sake
of comparison. The contrast between the simulations and the
experimental curve seems to reveal a systematic deviation such
that the simulated and RMSA profiles are actually somewhat too
steep. Indeed, when we allow $Z$ to be a fit parameter in the
one-component Yukawa system, keeping all the other parameters
equal, it turns out that $Z=94$ gives best agreement of the
one-component models with the experimental profile. It is tempting
to conclude, therefore, that $Z=76$ gives merely a best fit to the
experiment within PB theory given by Eqs.(\ref{PB1})-(\ref{PB3}),
which (to a large extent) ignores colloid-colloid correlations,
whereas inclusion of these correlations (as in the simulations of
the primitive model and that of the Yukawa system, and in the
RMSA-based theory) gives rise to a density profile that is
systematically steeper in comparison with the experiment.

\begin{figure}
\label{theories}
\includegraphics[width=70mm,angle=-90]{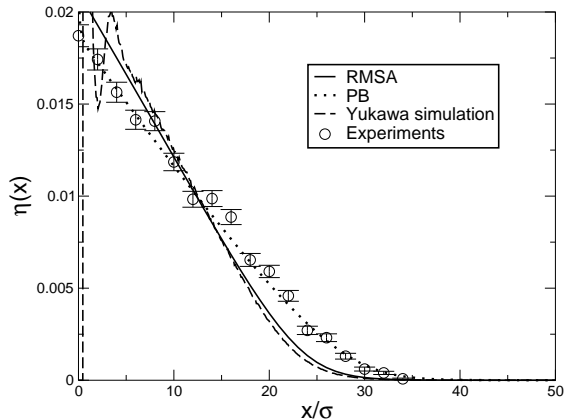}
\caption{Sedimentation profiles of the effective one-component Yukawa model calculated using both RMSA and standard MC simulations, compared with
the experimental measurements of Ref. \cite{Experiments} and the three-component PB theory, for $Z=76$ and all other parameters as in Fig.2. Note
the close agreement between the two Yukawa results, and their small but systematic deviation from the experiments and the PB theory.}
\end{figure}

\begin{figure}
\label{OC94}
\includegraphics[width=70mm,angle=-90]{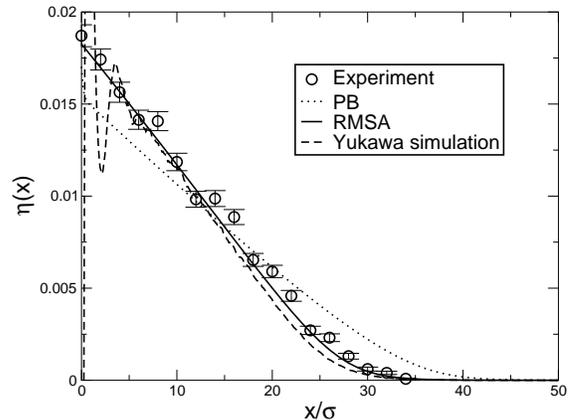}
\caption{As Fig.3, but with $Z=94$, such that the Yukawa model profiles (both RMSA and simulated) fit the experimental profile best. The PB
profile is clearly less accurate now.}
\end{figure}

In figure 4 we show the resulting sedimentation profiles based on
the Yukawa potential simulations and the RMSA closure for $Z=94$.
For comparison we also plot the multi component PB model for
$Z=94$ revealing a relatively poor agreement with the other
curves. In this case the PB approach clearly fails to reproduce
quantitatively the sedimentation profiles. On the other hand, the
agreement of the experimental profile with the effective
one-component models is good, except that as mentioned before, the
simulated profile exhibits much more structure close to the
hard-wall that represents the bottom of the sample in the
simulations ---this packing effect is not captured by the
local-density approximation that underlies the hydrostatic
equilibrium condition, and is not seen in the experiment because
the actual sample extends beyond the plotted $x$-range.

\begin{figure}
\label{allSim}
\includegraphics[width=70mm,angle=-90]{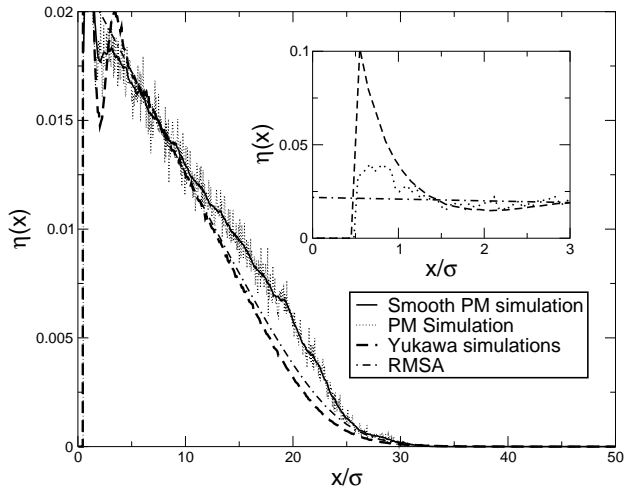}
\caption{Sedimentation density profiles as obtained from simulations of the Primitive Model in gravity and simulations of the Yukawa fluid
compared with the one-component RMSA model for colloidal charge $Z=76$ and all other parameters as in Fig.2. The difference between the profiles
can partly be attributed to the structural differences close to the bottom at $x=0$ as shown in the inset, where the simulations reveal hard-wall
induced structure that is not captured by the RMSA-based theory, and perhaps partly by slow equilibration and poor statistics in the simulations
due to the small number of colloids. }
\end{figure}

In Fig.5 we show sedimentation profiles as obtained by simulations of the primitive model and of the Yukawa model, compared with those by the RMSA
approach, all for $Z=76$. The agreement is perhaps a bit less quantitative than one would have expected. One of the reasons that the
 density in the primitive model is considerably higher in $x/\sigma\in(10,25)$ is due to the structure close to the hard wall at the
bottom near $x=0$, as shown in the inset of Fig.5, where the two Yukawa systems reveal a larger net adsorption than that of the primitive model,
albeit for different reasons: the simulated Yukawa system shows a strong peak at $x=\sigma/2$ while the RMSA-based profile continues to be nonzero
down to $x=0$. Given that we imposed that $\bar{\eta}$ is identical in all cases, there must also be a region in space where the density in the
primitive model exceeds the other two; the order of magnitude of the integrated differences over $x/\sigma\in(10,25)$ is indeed similar to the
negative of that over $x/\sigma\in(0,3)$. Another reason for these differences might be the poor statistics and slow equilibration of the
primitive model simulations. Recall that the present data are based on about one year of CPU time, so considerable extensions and more checks are
not easily obtained. This also prevented us from performing primitive model simulations for $Z=94$.

\section{Discussion}

The first observation from Fig. 3 should be the gross agreement between the experiments and all calculated and simulated profiles. In all cases we
have $\kappa\sigma=1.2$, Bjerrum length $\lambda_B=10.4$ nm, colloid diameter $\sigma=1.91\mu$m, and gravitational length $L=2.41\sigma$. The
colloidal charge is taken as $Z=76$, and the measured packing fractions are in the range $0<\eta(x)<0.02$, where $\eta(x)=(\pi/6)\sigma^3\rho(x)$.
This regime is such that $Z\rho(x)/2\rho_s<0.17$ for all $x$, i.e. even at the highest density the ion concentration is dominated by the reservoir
salt concentration $2\rho_s$, such that the screening constant is indeed essentially a constant independent of the height or density, as
implicitly assumed in equation (\ref{DLVO Potential}). A closer look, however, shows that even though the simulations and the RMSA result of the
Yukawa system are very close to each other (except at the bottom where packing effects affect the simulations), they both deviate systematically
from the experiment: the former two are too steep and have too low a density at higher altitudes. From the fact that the RMSA-based profile and
that of the Yukawa simulations are so close to each other, one could conclude that they are mutually consistent and both accurate, and that their
deviation from the experiment is mainly due to the present choice of $Z=76$, which was based on the fitting to the PB theory of
Eqs.(\ref{PB1})-({\ref{PB3}).  This fitting is not optimal due to the inadequacy of the present PB theory to account for correlations among the
different species in the system. In particular, PB theory overestimates the colloid density at high altitudes for a given value of $Z$. This
happens for charges $Z\gtrsim50$, whereas for smaller values of the colloid charge the agreement between the profiles obtained from the PB and
RMSA approaches is excellent, as far as an effective (smaller) value for the charge is used in the PB approach as discussed in detail below. For
the parameters of present interest, fitting the experimental density profile to that of the Yukawa system treated within the RMSA closure, we
concluded that $Z=94$ gives the best fit.

Given that a Yukawa system with a colloidal charge $Z=Z_{RMSA}=94$ is best described within PB theory by a colloidal charge $Z=Z_{PB}=76$, for the
present system parameters, it is interesting to investigate the relation between $Z_{RMSA}$ and $Z_{PB}$ for other values of the colloid charge.
In figure 6 we plot the ratio $Z_{PB}/Z_{RMSA}$ for $0<Z_{RMSA}<200$, for three screening constants while all the other parameters are left
unchanged. Figure 7 shows the corresponding density profiles for $\kappa\sigma=1.2$ and $Z_{RMSA}=200$, 94, and 40. One can conclude that for
$Z_{RMSA}\lesssim50$, PB theory reproduces the RMSA-Yukawa sedimentation profiles very accurately provided the colloidal charge $Z_{PB}$ is
reduced by up to $20\%$ of $Z_{RMSA}$. Note that the average density here is low enough that the profile in the limit of $Z_{RMSA}\rightarrow 0$
becomes essentially barometric; at higher average packing fractions one also expects $Z_{PB}/Z_{RMSA}\neq 1$ in this limit, due to hard-core
effects that are not accounted for properly in the PB theory. The required reduction of $Z_{PB}/Z_{RMSA}$ with increasing $Z_{RMSA}>50$ exceeds
$20\%$, and, in addition, the quality of the best-fitting PB profile (quantified by the mean-squared deviation) becomes slightly less satisfactory
as is reflected by the increase of the error bars in figure 6 with increasing $Z_{RMSA}$, this is also shown in figure 7.

\begin{figure}
\label{Zeff}
\includegraphics[width=70mm,angle=-90]{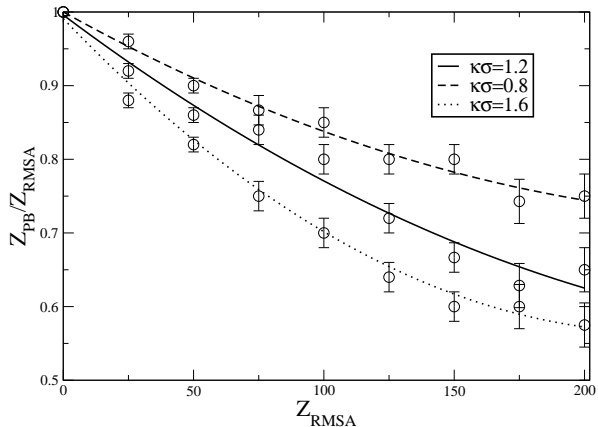}
\begin{center}
\caption{Ratio of the best-fitting colloidal charge $Z=Z_{PB}$ (see text) and that of the RMSA charge $Z_{RMSA}$, as a function
$Z_{RMSA}\in(0,200)$, with $\sigma$, $\lambda_B$, $L$, $\bar{\eta}$, and $H$ as in Fig.1, for various screening parameters $\kappa\sigma$. Note
that PB theory is increasingly better for lower colloidal charges and lower salt concentrations. The lines are mere guides to the eye.}
\end{center}
\end{figure}

\begin{figure}
\label{Zeff2}
\includegraphics[width=70mm,angle=-90]{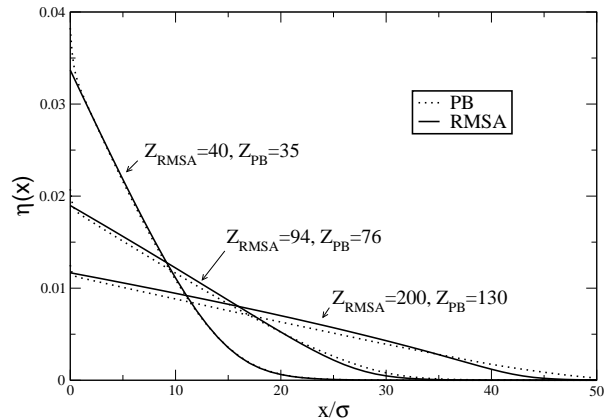}
\begin{center}
\caption{Sedimentation profiles obtained from RMSA theory compared to those of PB theory using the best-fitting $Z_{PB}$ from Fig.6, for
$\kappa\sigma=1.2$ and all the other parameters as in Fig.6. Note that the quality of the fit detoriates slightly with increasing $Z_{RMSA}$. }
\end{center}
\end{figure}

The difference between $Z_{RMSA}$ and $Z_{PB}$ in the present
system is of course considerable and significant, but not
qualitative. The seemingly different mechanisms that underly the
lifting of the colloids to higher altitudes than given by the
barometric distribution, as predicted by the three- and
one-component theory, should therefore be actually equivalent: the
self-consistent electric field that is generated by a net
charge-imbalance at the boundaries of the 3-component system such
that the colloids are pushed upwards is merely another way of
describing a pairwise screened-Coulomb repulsion that pushes the
colloids apart to higher altitudes in a one-component model. This
is in line with conclusions in Ref.\cite{belloni}.

We have attributed the difference between the best fit for $Z$
based on Poisson-Boltzmann theory and the other three methods
(simulations of the primitive model, and simulations and RMSA
theory of the one-component Yukawa system) to the poor account of
correlations in the Poisson-Boltzmann theory. In principle,
however, there could be other sources that cause such a
difference, e.g. charge renormalization and hard-core exclusion
effects for the screening ions. Charge renormalization due to
nonlinear screening effects \cite{Alexander} is, however, {\em
not} a candidate here to explain the difference for at least two
reasons: (i) The actual bare charge is usually larger than the
renormalized charge that appears in the prefactor of the
screened-Coulomb interactions, whereas here the former seems to be
smaller. (ii) The present parameters here are such that
$Z\lambda_B/a<1$, whereas charge renormalization is only
substantial if this dimensionless combination exceeds about 5 or
so \cite{Alexander,Bas,Trizac}. A mechanism whereby the effective
colloidal charge is increased was discussed in Refs. \cite{more
Belloni,Krampost}, and is based on  the hard-core exclusion of the
screening ions at sufficiently high colloid packing fractions: the
screening is therefore less effective, which appears as an
increase of the effective colloidal charge. However, applying the
analysis of Ref.\cite{more Belloni} to the present case gives only
marginally larger values of the effective charge, by less than 1\%
at the highest density $\eta\simeq 0.02$. In other words, it
appears that this effect cannot explain the difference between
$Z=76$ and $Z=94$, leaving the poor account of colloid-colloid
correlations in the Poisson-Boltzmann theory as the most plausible
source of the difference.

It is interesting to inquire whether the one- and three-component models would also produce essentially the same sedimentation profiles for other
sets of parameters than considered here, and whether the hydrostatic equilibrium condition (\ref{hydro}) and the Poisson-Boltzmann theory
(\ref{PB1})-(\ref{PB3}) for the one-component and the three-component case, respectively, produce reliable profiles in all circumstances. In order
to answer these questions we consider the regimes of extremely low and extremely high salt concentrations. First, the hamiltonian
(\ref{Hamiltonian One}) for the effective one-component system is pairwise additive, which is expected to be a good approximation for the present
parameter set where $\kappa\sigma>1$ and $Z\rho/2\rho_s<1$, i.e. the range of the interactions is smaller than the size of the particles and the
background electrolyte concentration dominates the counterion concentration. At lower salt concentrations, such that $\kappa\sigma\ll1$ and/or
$Z\rho/2\rho_s>1$, one would expect effective many-body interactions to become relevant \cite{Bas,Russ}, such that (\ref{Hamiltonian One}) is not
necessarily a reliable effective hamiltonian anymore. In such an extremely low-salt regime, which is realized in salt free systems, the
Poisson-Boltzmann theory proved to be quantitatively accurate, at least in comparison with simulations \cite{MC Simulations, Dijkstra-Jos} at low
Coulomb couplings. It is interesting to see if the pairwise one-component description is capable of describing the density profiles in this case.
We wish to stress here that the possible break-down of the pairwise screened-Coulomb picture does {\em not} imply that the system can no longer be
seen as a one-component system in hydrostatic equilibrium as described by (\ref{hydro}) with a compressibility that follows from the Kirkwood-Buff
relation $S(0)=\chi_T$: these relations remain valid (the former only within the local density approximation, but given the long screening length
in the extremely low-salt regime this approximation is probably accurate). The breakdown would "merely" imply that it is not obvious how to
calculate the compressibility or the structure factor without detailed knowledge of the effective hamiltonian. Second, let us consider the
opposite high-salt regime such that $\kappa\sigma\gg1$ and $Z\rho\ll 2\rho_s$. In this regime the electrostatic interactions are completely
screened over distances much smaller than the colloidal diameter, such that the effective one-component system is essentially a (pairwise)
hard-sphere system (for water at room temperature at least, where ion-ion correlations are not all that important). In this regime the
one-component description based on (\ref{hydro}) is far superior over the PB theory of Eqs. (\ref{PB1})-(\ref{PB3}). This is directly seen by
regarding the $Z=0$ limit of equations (\ref{PB1})-(\ref{PB3}), which reduce to $\phi(x)=0$ and $\rho(x)=\rho_0\exp[-x/L]$, i.e. the sedimentation
profiles become barometric; the hard-core correlations are left-out completely from this theory. By contrast, the RMSA closure is, in this
hard-core limit, equivalent to the Percus-Yevick closure, and in combination with (\ref{hydro}) the density profiles of hard-sphere sedimentation
equilibrium are well-described \cite{Piazza,Rutgers}. Moreover, also in the present regime with $\kappa\sigma\simeq 1$ the one-component theory
performs better. We are currently working on the formulation of a theory that is able to describe sedimentation density profiles in both the
high-salt and the low-salt regime on the same footing.

\section{Summary and Conclusions}
In this paper we have studied sedimentation equilibrium of charge-stabilized colloids at non-zero salt concentration. We compared experimental
results with theoretical and simulated profiles obtained on the basis of two models. On one hand a multi-component model of point-like colloids,
cations, and anions interacting with bare Coulomb potentials. For this model we considered a Poisson-Boltzmann theory of the three-component
mixture and performed MC simulations using 12 colloids and a total of about 26132 particles to guarantee the electroneutrality of the system. On
the other hand we considered an effective one-component model of colloids interacting by an effective screened-Coulomb potential. For this model
we employed a theory based on hydrostatic equilibrium, where the isothermal compressibility is given by the Kirkwood-Buff relation as obtained
from the solution of the Ornstein-Zernike equation with the rescaled mean spherical approximation (RMSA) closure for the screened-Coulomb
potential. For the effective one-component Yukawa model, we also performed simulations of sedimentation profiles.

The sedimentation profiles obtained from the one-component RMSA theory, simulations of the Yukawa system, and simulations of the primitive model
are essentially consistent among themselves but differ from the results of the Poisson Boltzmann theory. The PB theory shows good agreement with
the experiments only because the numerical value of the charge was estimated as to give the best fitting according to this theory. In fact, we
have seen that PB theory actually overestimates the colloid density at high altitudes compared to the corresponding Yukawa system, for identical
values of $Z$, at the parameters of interest here. Agreement between PB theory and Yukawa systems can be obtained by reducing the colloidal charge
in the PB theory compared to that of the Yukawa model. For small values of the colloid charge, $Z\lesssim 50$ or so, the agreement between the
resulting profiles obtained from the PB and RMSA approach is truly excellent, for larger charges up to say $Z=200$ the agreement is still good
though somewhat less quantitative as regards the functional form. The CPU time required for the simulation of the multi-component primitive model
and the effective one-component Yukawa model varies between about one year in the former case and one hour in the latter. This shows that theory
and simulations of sedimentation profiles on the basis of the effective one-component potential and the Poisson-Boltzmann theory (possibly with a
reduced effective charge when colloid-colloid correlations are important) are considerably more efficient than primitive model simulations. In
spite of the fact that we have only considered a particular case study, this is presumably true in general, with the Yukawa model probably more
accurate when $\kappa\sigma\gtrsim 1$ while PB theory could be more accurate or efficient  when $\kappa\sigma\ll 1$. This will be investigated in
more detail in a future study.

\section{Acknowledgments}
\begin{acknowledgments}
It is a pleasure to thank Alfons van Blaaderen  and Paddy Royall
for collaborations and sharing their sedimentation profiles data
with us. This work is part of the research program of the "Stichting voor Fundamenteel Onderzoek der Materie (FOM)", which is financially supported by the
"Nederlandse Organisatie voor Wetenschappelijk Onderzoek (NWO)". NWO-CW is acknowledged for the TOP-CW funding.
\end{acknowledgments}

\section{}
\subsection{}
\subsubsection{}

\end{document}